\documentclass[aps,prd,amsmath,floats,floatfix, twocolumn, superscriptaddress,nofootinbib,showpacs]{revtex4-1}

\usepackage[colorlinks, pdfborder={0 0 0}, plainpages=false]{hyperref}
\usepackage{graphicx}
\usepackage{xspace}
\usepackage[usenames,dvipsnames]{color}
\usepackage{amssymb}
\usepackage{xcolor}
\usepackage[normalem]{ulem} 
\usepackage{array}

\definecolor{mpl_red}{HTML}{D62728}

\newcommand\be{\begin{equation}}
\newcommand\ba{\begin{eqnarray}}
\newcommand\ee{\end{equation}}
\newcommand\ea{\end{eqnarray}}
\newcommand{\nn}{\nonumber}

\newcommand{\s}{{\mbox{\tiny SS}}} 
 
\newcommand{\RR}{{\mbox{\tiny RR}}}

\begin{document}

\title{
Parameter control for eccentric, precessing binary black hole simulations with SpEC
}

\author{Taylor Knapp}
\email{tknapp@caltech.edu}
\affiliation{Department of Physics, California Institute of Technology, Pasadena, California 91125, USA}
\affiliation{LIGO Laboratory, California Institute of Technology, Pasadena, California 91125, USA}

\author{Katerina Chatziioannou} 
\email{kchatziioannou@caltech.edu}
\affiliation{Department of Physics, California Institute of Technology, Pasadena, California 91125, USA}
\affiliation{LIGO Laboratory, California Institute of Technology, Pasadena, California 91125, USA}

\author{Harald Pfeiffer}
\email{harald.pfeiffer@aei.mpg.de}
\affiliation{Max Planck Institute for Gravitational
    Physics (Albert Einstein Institute), Am M\"{u}hlenberg 1, 14476
    Potsdam-Golm, Germany}

\author{Mark A. Scheel} 
\email{scheel@tapir.caltech.edu}
\affiliation{Department of Physics, California Institute of Technology, Pasadena, California 91125, USA}

\author{Lawrence E. Kidder} 
\email{kidder@astro.cornell.edu}
\affiliation{Cornell Center for Astrophysics and Planetary Science, Cornell University, Ithaca, New York 14853, USA}

\date{\today}

\begin{abstract}
Numerical relativity simulations of merging black holes provide the most accurate description of the binary dynamics and the emitted gravitational wave signal.
However, practical considerations such as imperfect initial data and initial parameters mean that achieving target parameters, such as the orbital eccentricity or the black hole spin directions, at the beginning of the usable part of the simulation is challenging.
In this paper, we devise a method to produce simulations with specific target parameters, namely the Keplerian orbital parameters—eccentricity, semimajor axis, mean anomaly—and the black hole spin vectors using SpEC.
The method is an extension of the current process for achieving vanishing eccentricity and it is based on a parameter control loop that iteratively numerically evolves the system, fits the orbit with analytical post-Newtonian equations, and calculates updated input parameters.
Through SpEC numerical simulations, we demonstrate $\lesssim 10^{-3}$ and ${\cal{O}}(\rm degree)$ convergence for the orbital eccentricity and the spin directions respectively in $\leq7$ iterations.
These tests extend to binaries with mass ratios $q \leq 3$, eccentricities $e \leq 0.65$, and spin magnitudes $|\chi | \leq 0.75$.
Our method for controlling the orbital and spin parameters of numerical simulations can be used to produce targeted simulations in sparsely covered regions of the parameter space or study the dynamics of relativistic binaries.
\end{abstract}


\maketitle

\section{Introduction}

With no known analytic solutions, the two-body problem in general relativity can only be solved exactly with full numerical relativity (NR) simulations~\cite{Baumgarte:2002jm}. 
Such simulations have numerous practical applications in calibrating waveform models~\cite{Pratten:2020ceb,Ramos-Buades:2023ehm} or serving as the basis of surrogate models~\cite{Varma:2019csw} that are used in gravitational wave data analysis, e.g.,~\cite{LISAConsortiumWaveformWorkingGroup:2023arg,LIGOScientific:2016aoc,LIGOScientific:2020iuh}.
Moreover, they provide solutions to the full spacetime and elucidate the properties of black holes (BHs) and general relativistic dynamics.
Modern NR codes such as the spectral Einstein code (SpEC) have produced thousands of simulations of coalescing BHs~\cite{Boyle:2019kee}.

NR solves an initial value problem where initial conditions are evolved forward in time. 
For binary BH (BBH) systems, one can freely specify
the ratio of the BH masses, the initial BH spin angular momenta (magnitude and direction), and the initial coordinate positions and velocities of the BHs.
For puncture data~\cite{Brandt:1997tf} these are indeed directly the free parameters.  For quasiequilibrium conformal thin sandwich data~\cite{Pfeiffer:2002iy,Cook:2004kt,Caudill:2006hw} ---which is the type of initial-data considered here, as it provides initial lapse and shift for evolutions and allows for nearly extremal black holes~\cite{Lovelace:2008tw}--- the positions and velocities of the BHs are instead encoded by the binary's orbital velocity, the BH separation, and the BH relative radial velocity.
Evolving the initial data forward in time yields  the full spacetime dynamics, BH properties, and emitted gravitational wave signal as a function of time.

However, in practice the initial data for a BBH simulation do not correspond to a snapshot of a binary that has been inspiraling since the infinite past.  
This is because the initial data do not include the correct gravitational radiation that was emitted in the past during the infinitely long inspiral, nor do they include the appropriate tidal distortions of the BHs.
As a result, when an NR simulation begins to evolve, initial transients occur before the BBH relaxes into a quasiequilibrium state.  
The gravitational radiation emitted during this process is known as ''junk radiation'' 
Junk radiation, if it is resolved by the simulation, is a physical solution of the Einstein equations, but is not the solution of interest.  
There have been multiple attempts to reduce the amount of junk radiation in NR simulations~\cite{Alvi:1999cw,Yunes:2006iw,JohnsonMcDaniel:2009dq,Kelly:2009js,Reifenberger:2012yg,Tichy:2016vmv,Lovelace:2008hd,Varma:2018sqd, Etienne2024improvedmovingpuncture}, but currently the standard practice is to simply discard the first few orbits until the junk radiation has decayed away~\cite{Boyle:2019kee,pretto2024automateddeterminationendtime}.

A further complication arises in selecting initial parameters.
In Newtonian gravitation, it is straightforward to choose the initial positions and velocities of two point particles so that the resulting orbit has a desired semimajor axis $a$, eccentricity $e$, and mean anomaly $\ell$ at some reference time. 
But in general relativity with no known closed-form or ordinary differential equation solution, achieving a BBH orbit with desired parameters is not straightforward.

Among orbital parameters, the eccentricity is particularly astrophysically relevant. 
Since gravitational radiation reaction efficiently circularizes BBH orbits~\cite{PhysRev.136.B1224}, most simulations have targeted vanishing eccentricities.
Motivated by this, the SpEC workflow includes an initial ``eccentricity removal" stage that tunes the input parameters, i.e., the angular velocity, separation, and relative radial velocity, to reduce eccentricity below a predetermined threshold~\cite{Buonanno:2010yk, habib2024eccentricityreductionquasicircularbinary}. 
An iterative scheme makes initial guesses for the input parameters, performs an NR simulation for a few orbits, measures the eccentricity, and then chooses updated input parameters for the next iteration.
However, recent hints of nonzero eccentricity in select observed signals~\cite{Romero-Shaw:2021ual,Romero-Shaw:2022xko,Ramos-Buades:2023yhy,Gupte:2024jfe} and an emphasis of eccentricity as a tool to study the BBH formation history~\cite{Zevin:2021rtf} have reinvigorated interest in eccentric BBHs~\cite{Romero-Shaw:2022fbf,Ramos-Buades:2021adz,Nagar:2024dzj}.

Another astrophysically relevant property is the BH spin, which determines the length
and morphology of the signal~\cite{KAGRA:2021duu} and whose value also carries information about the BBH formation history.
The NR spins are set as part of the initial data construction, but the spin directions precess if the spins are misaligned with the orbital angular momentum~\cite{Apostolatos1994}.  
If the first few orbits of the simulation are discarded because of junk radiation, then the useable part of the simulation will begin with slightly different spin directions, whose values are difficult to predict because they depend on how long the junk radiation lasts.
Thus, it is difficult to precisely control spin directions at the beginning of the usable part, i.e., after junk radiation, of a precessing NR simulation.


In this paper we extend the eccentricity removal procedure of Ref.~\cite{Buonanno:2010yk} to nonzero eccentricity; our method also specifies the spin directions at some chosen time different than $t=0$.
After introducing the problem setup, in Sec.~\ref{sec:methods} we describe the method.
Given the target parameters of the simulation we would like to evolve, i.e., target Keplerian parameters and spins, we iteratively evolve the system, extract the binary parameters at the usable part of the data, and adjust the input parameters.
Each evolution proceeds for 3-5 orbits, at the end of which we fit the numerical data with a post-Newtonian model and extract the Keplerian parameters. 
We then use standard root-finding techniques to compute new input parameters that when evolved will result in a simulation that is closer to the target parameters.

We validate this method in Secs.~\ref{sec:aligned} and~\ref{sec:precessing} for systems with aligned and precessing spins respectively.
Selecting a threshold of achieving the target eccentricity to within $7\times10^{-4}$ motivated by the zero-eccentricity limit, we show that $\leq7$ iterations are sufficient for binaries with mass ratio $q \leq 3$, eccentricity $e \leq 0.65$, and spin magnitude $\chi \leq 0.75$.
Simultaneously, the spin directions are fixed with ${\cal{O}}(\rm degree)$ accuracy relative to their target directions. We conclude in Sec.~\ref{sec:conclusions}.

\section{Initial conditions for Numerical Relativity}
\label{sec:methods}

Consider a binary of two BHs with masses $m_A$, with $A \in \{1,2\}$, total mass $M\equiv m_1+m_2$ , mass ratio $q\equiv m_1/m_2 \geq 1$, and symmetric mass ratio $\eta=q/(1+q)^2$. The initial dimensionless spins of the BHs are denoted $\vec{\chi}_A$. 
To simulate such a binary, SpEC requires initial data parameters $\vec{\theta}_{\rm ID} = \{\Omega_0,v_{r,0},D_0\}$ where $\Omega_0$ is the orbital angular frequency, $v_{r,0}$ is the coordinate relative radial velocity, and $D_0$ is the coordinate BH separation at simulation time $t=0$. 
Other SpEC papers adopt the convention $\dot a_0=v_{r,0}/D_0$ instead of $v_{r,0}$, e.g.~\cite{Buonanno:2010yk,Ossokine:2015yla}.  
We use $v_{r,0}$ here and reserve the letter $a$ for the semimajor axis of elliptical orbits.

As the evolution proceeds, the coordinate trajectories are described by the BH positions $\vec{x}_A(t)$.  
If the orbit were Newtonian, the trajectory could be equivalently described by a Keplerian parametrization in terms of orbital elements $\vec{\theta}_{\rm orb} = \{a,e,\ell\}$, where $a$ is the semimajor axis, $e$ is the eccentricity, and $\ell$ is the mean anomaly at some reference time. 
The actual trajectories are however not Newtonian, but rather post-Newtonian (PN) with an increasing number of relevant corrections as the BHs approach merger~\cite{Boyle:2007ft}. 
As we consider a small portion of the whole trajectory far from merger, namely the first few orbital periods, and we still approximately parametrize the orbit by $\vec{\theta}_{\rm orb}$. 
We describe this parametrization and correspondence between $\vec{\theta}_{\rm orb}$ and $\vec{\theta}_{\rm ID}$ in Sec.~\ref{sec:fittingmodel}.

\begin{figure}[] 
\includegraphics[width=.95\columnwidth,clip=true]{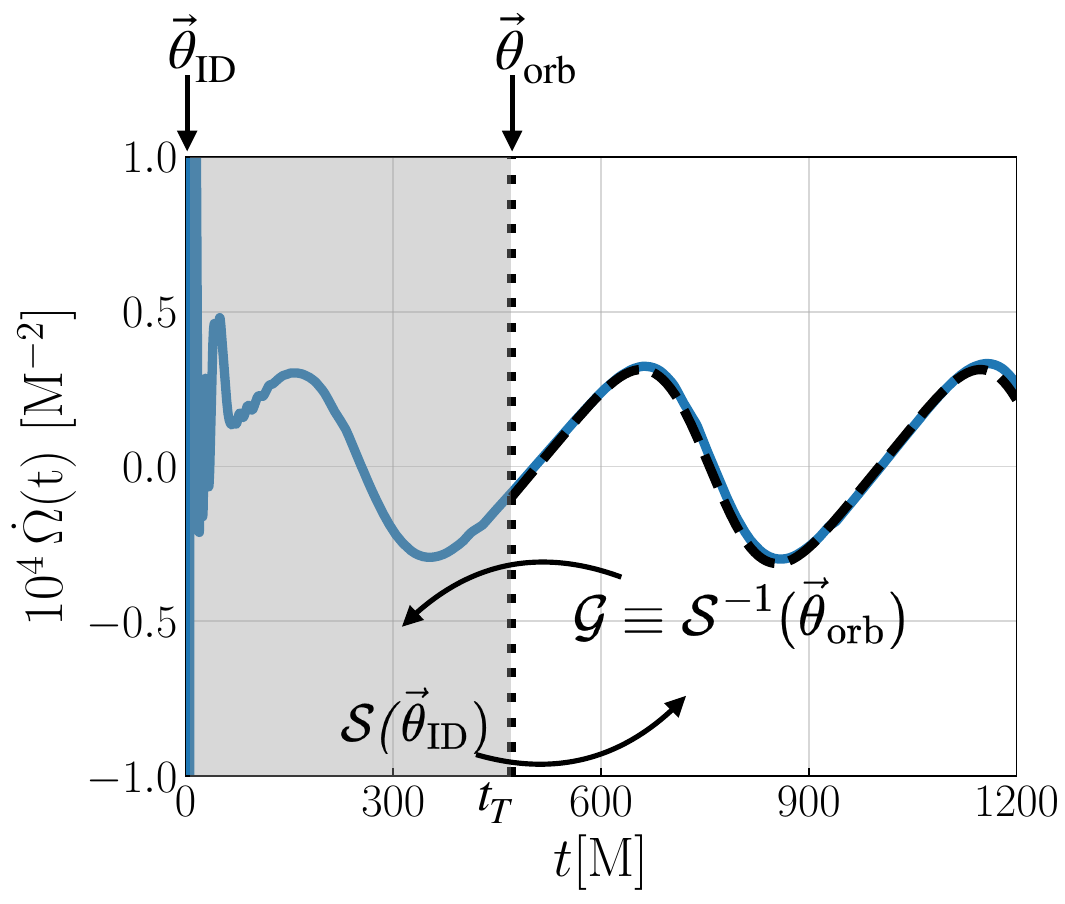}\\
\caption{Schematic representation of the challenges in determining the initial conditions of numerical relativity simulations. 
We use an example simulation and plot the derivative of the orbital angular frequency as a function of time, $\dot{\Omega}(t)$, in blue. 
The gray region from $t = 0$ to $t = t_T$ (dotted vertical line) is the window of junk radiation. We overlay a post-Newtonian fitted solution (dashed) using Eq.~\eqref{eqn:fittingmodel}. 
Our root-finding method is represented by $\mathcal{G}$ and $\cal{S}$ as introduced in Sec.~\ref{sec:root}. 
The junk radiation contributes a significant nonlinear effect in the first $\sim 100$ M of the simulation.}
\label{fig:method}
\end{figure}

The goal is to devise a method of choosing appropriate input values of the initial data parameters $\vec{\theta}_{{\rm ID},T}$ and the initial spins $\vec{\chi}_{A,T}(t=0)$ so that we obtain simulations with target orbital parameters $\vec{\theta}_{\rm orb}{}_{,T}$ and spins $\vec{\chi}_{A,T}(t=t_T)$ at some time $t_T$.
Throughout the paper, we use the subscript $T$ to denote the ``target" parameters, including both the orbital parameters $\vec{\theta}_{\rm orb}{}_{,T}$ and the corresponding initial data parameters $\vec{\theta}_{{\rm ID},T}$ that when evolved will give an orbit with $\vec{\theta}_{\rm orb}{}_{,T}$ (and similarly for the spins).
Figure~\ref{fig:method} presents the problem setup in schematic form. 
Input parameters $\vec{\theta}_{\rm ID}$ and initial spins $\vec{\chi}_{A}(t=0)$ are set at the initial time $t = 0$ and the system is numerically evolved (blue).
The usable portion of the numerical data starts after the end of the junk radiation phase (gray band), denoted at $t_T$ (black dotted vertical line).
At that time, the spins $\tilde{\chi}_{A}(t=t_T)$ are read off the numerical data, while $\vec{\theta}_{\rm orb}$ is obtained as described in Sec.~\ref{sec:fittingmodel}.

When specifying the target spins $\vec{\chi}_{A,T}(t=t_T)$ at the target
time $t_T$, a suitable reference frame needs to be selected.  By
convention, SpEC simulations begin at $t=0$ with a frame whose
$z$-axis is aligned with the orbital angular momentum and the BHs are
on the $x$-axis, with the larger BH on the positive $x$-axis.
During the simulation, the BHs move in these \textit{inertial coordinates}, and the simulation outputs the BH \textit{inertial frame components} of the spins. At time $t_T$---at which we want to set the BH spins---the BHs will in general
  not be on the inertial coordinate x-axis, and will in general not
  move in the \textit{xy}-plane of the inertial coordinates.  In order to specify target
  spins independent of the precise inertial frame position of the black holes, we introduce the \textit{BH frame}, which is coorbiting in the sense that its \textit{x}-axis is always pointing from BH 1 to BH 2, and its \textit{z}-axis is orthogonal to the instantaneous orbital plane.  (At $t=0$, the BH frame coincides with the inertial frame). 
For clarity, we will denote spins in the BH frame with a vector, e.g., the target spin is $\vec{\chi}_{A,T}(t=t_T)$, and spins in the inertial frame (which is the frame the NR data express them in) with a tilde.

Two challenges arise when attempting to map from some chosen initial parameters $\vec{\theta}_{\rm ID}$ and $\vec{\chi}_{A}(t=0)$ to orbital parameters   $\vec{\theta}_{\rm orb}$ and $\vec{\chi}_{A}$ at time $t=t_T$:
\begin{enumerate}
    \item Due to junk radiation, we have to set $t_T\gtrsim100\,$M. 
    During that time, spin-precession changes the BH spin directions in the inertial simulation frame. This change happens on the (slow) precession timescale.
    In the co-orbiting BH frame where
      we define the target spins,
    the spins evolve on the much faster orbital timescale. 
    To minimize impact of this fast timescale, we select $t_T$ to correspond to an integer number of orbits, after which the BH frame has roughly returned to its initial position. Here, we choose $t_T$ to be one orbit. This choice is explained further in Sec.~\ref{reference_time}.
  \item The mapping between $\vec{\theta}_{\rm ID}$ and $\vec{\theta}_{\rm orb}$ is not straightforward under full general relativistic dynamics, so that initial data parameters $\vec{\theta}_{{\rm ID},T}$ that yield the desired target orbital parameters $\vec\theta_{{\rm orb},T}$ are initially not known.
    Therefore, we construct an iterative update approach to obtain $\vec{\theta}_{\rm orb,\,T}$ through small updates to $\vec{\theta}_{\rm ID}$.
    The iterative scheme is formalized and explained in depth in the rest of this section.
      
\end{enumerate}
For reference in the remainder of this text, we supply Table~\ref{tab:vars} with definitions of parameter notations. 

\begin{table}
\renewcommand{\arraystretch}{1.5}
\begin{tabular}{|c|p{7cm}|}
     \hline
     Variable & Description  \\ 
     \hline
     $\vec{\chi}_{A,T} (t=t_T)$       & Target spin vector for BH A at $t=t_T$ in the coorbiting BH frame\\
     \hline
     $\vec{\chi}_{A,T} (t=0)$       & Initial spin vector for BH A that when evolved will result in $\vec{\chi}_{A,T} (t=t_T)$\\
     \hline
     $\vec{\chi}_{A}^{\,(i)} (t=0)$& Initial spin vector for BH A for the $i$th trial simulation\\
     \hline
     $\tilde{\chi}_{A}^{\,(i)}(t=t_T)$& Spin vector for BH A at $t=t_T$ in the inertial frame for the $i$th trial simulation, read off the NR data\\
     \hline
     $\vec{\theta}_{\rm orb}{}_{,T}$   & Target orbital parameters at $t=t_T$\\
     \hline
     $\vec{\theta}_{{\rm ID},T}$    & Target initial data parameters that when evolved will result in $\vec{\theta}_{\rm orb}{}_{,T}$\\
     \hline
     $\vec{\theta}_{\rm ID}^{\,(i)}$& Initial data parameters for the $i$th trial simulation\\
     \hline
     $\vec{\theta}_{\rm orb}^{\,(i)}$& Orbital parameters for the $i$th trial simulation\\
     \hline
\end{tabular}

\caption{A reference table for the initial parameters, orbital parameters, and spin utilized in this work. An arrow or tilde over each spin parameter refers to the BH or inertial frame, respectfully, that it is defined in. \label{tab:vars} }
\end{table}

\subsection{The inverse problem and root-finding}
\label{sec:root}

The map between the input parameters $\vec{\theta}_{\rm ID}$ at $t=0$ and the orbital parameters $\vec{\theta}_{\rm orb}$ at a later time $t_T$ is written as an operator, ${\cal{S}}$ which includes
\begin{enumerate}
    \item The ``forward" mapping from $\vec{\theta}_{\rm ID}$ at $t=0$ to the numerical data at later time $t_T$; this is the SpEC evolution.
    \item The extraction of $\vec{\theta}_{\rm orb}$ that parametrizes the numerical data.
\end{enumerate}
Collectively, these steps are denoted as, ${\cal{S}}(\vec{\theta}_{\rm ID})$ in Fig.~\ref{fig:method}. The inverse mapping is denoted by ${\cal{G}}$ in Fig.~\ref{fig:method}. While ${\cal{S}}(\vec{\theta}_{\rm ID})$ can be fully calculated with SpEC, $\mathcal{G}$ is unknown, which we will elaborate upon later.

For some target orbital parameters, $\vec{\theta}_{\rm orb}{}_{,T}$, we seek $\vec{\theta}_{\rm ID,\, T}$ such that
\begin{equation}\label{eq:defS}
{\cal{S}}(\vec{\theta}_{\rm ID,\,T}) = \vec{\theta}_{\rm orb}{}_{,T}\,.
\end{equation}
Suppose we begin with an initial data guess $\vec{\theta}_{\rm ID}^{\,(i)}$ that yields orbital parameters $\vec{\theta}_{\rm orb}^{\,(i)}$. 
The superscript here denotes the $i$th trial evolution as part of an iterative scheme.
If we update $\vec{\theta}_{\rm ID}^{\,(i)}$ to $\vec{\theta}_{\rm ID}^{\,(i+1)}=\vec{\theta}_{\rm ID}^{\,(i)}+\delta \vec{\theta}_{\rm ID}$, a Taylor expansion yields
\begin{equation} \label{eq:S_approx}
{\cal{S}}\left(\vec{\theta}_{\rm ID}^{\,(i)}+\delta \vec{\theta}_{\rm ID}\right) = {\cal{S}}\left(\vec{\theta}_{\rm ID}^{\,(i)}\right)+\frac{\delta{\cal{S}}}{\delta \vec{\theta}_{\rm ID}} \delta \vec{\theta}_{\rm ID}\,. 
\end{equation}
 Requiring that the updated evolution brings us to the target parameters ${\cal{S}}(\vec{\theta}_{\rm ID}^{\,(i)}+\delta \vec{\theta}_{\rm ID}) = \vec{\theta}_{\rm orb}{}_{,T}$ yields
\begin{align}
\delta \vec{\theta}_{\rm ID} &= \left[ \frac{\delta{\cal{S}}}{\delta \vec{\theta}_{\rm ID}}\right]^{-1}\left[\vec{\theta}_{\rm orb}{}_{,T}-{\cal{S}}\left(\vec{\theta}_{\rm ID}^{\,(i)}\right)\right]\nn
\\
&= \frac{\delta{\cal{G}}}{\delta \vec{\theta}_{\rm orb}} \left[\vec{\theta}_{\rm orb}{}_{,T}-{\cal{S}}\left(\vec{\theta}_{\rm ID}^{\,(i)}\right)\right]\nn
\\
& \approx {\cal{G}}\left(\vec{\theta}_{\rm orb}{}_{,T}\right) - {\cal{G}}\left[{\cal{S}}\left(\vec{\theta}_{\rm ID}^{\,(i)}\right)\right]\nonumber
\\
& \approx {\cal{G}}\left(\vec{\theta}_{\rm orb}{}_{,T}\right) - \vec{\theta}_{\rm ID}^{\,(i)}\,,\label{eq:deltax_NR}
\end{align}
where ${\cal{G}} \equiv {\cal{S}}^{-1}$ is the inverse operator. 
To go from the first to the second line we have used the property of inverse functions 
\begin{equation}
\left[ \frac{\delta{\cal{S}}}{\delta \vec{\theta}_{\rm ID}}\right]^{-1}=\frac{\delta{\cal{G}}}{\delta \vec{\theta}_{\rm orb}}\,,
\end{equation}
while for the third line, we use the definition of the directional derivative, assuming that $\delta\vec\theta_{\rm ID}$ is small.
Equation~\eqref{eq:deltax_NR} provides a way to compute $\delta \vec{\theta}_{\rm ID} $ such that $\vec{\theta}_{\rm ID}^{\,(i+1)} = \vec{\theta}_{\rm ID}^{\,(i)}+\delta \vec{\theta}_{\rm ID}$ brings us closer to the target orbital parameters $\vec{\theta}_{\rm orb}{}_{,T}$.  

If we knew the operator $\cal G$ exactly, then the update would converge in one iteration using the procedure above.  Unfortunately, the exact $\cal G$ is unknown as it corresponds to the inverse operation of a full numerical simulation.  Therefore, the primary purpose of this paper is to derive suitable approximations to $\cal G$ such that the iterative procedure converges in a reasonable number of iterations. In Sec.~\ref{sec:iterations} we describe the iterative process, while in Sec.~\ref{sec:pn} we use post-Newtonian Keplerian equations to approximate the map between $\vec{\theta}_{\rm orb}$ and $\vec{\theta}_{\rm ID}$.  

\subsection{Parameter control loop}
\label{sec:iterations}

\begin{figure}[] \label{fig:eccflowchart}
\includegraphics[width=\columnwidth,clip=true,trim=6 6 0 8]{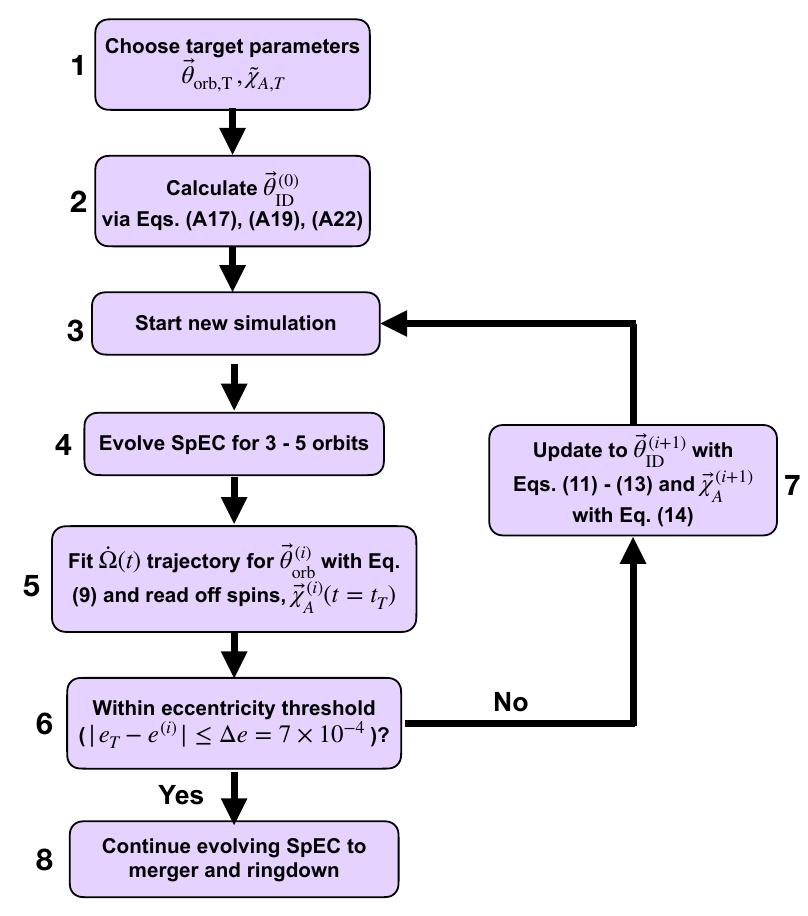}\\
\caption{ \label{fig:eccflowchart} Flow chart for the SpEC eccentricity and spin control loop, an extension of the zero-eccentricity loop of Ref.~\cite{Buonanno:2010yk}.}
\end{figure}
As will be further discussed in Sec.~\ref{sec:pn}, computing $\cal G$ includes a number of approximations. 
We therefore introduce a parameter control loop that iteratively performs trial simulations, extracts their parameters at $t_T$, updates the initial parameters, and starts new simulations until achieving a predefined tolerance threshold. 

The full workflow is presented in Fig.~\ref{fig:eccflowchart}.
\begin{enumerate}
\item We begin with a user-selected set of target orbital parameters $\vec{\theta}_{\rm orb}{}_{,T}$ and spins $\vec{\chi}_{A,T}$ (box 1).
\item We calculate the SpEC input parameters for the $i=0$ iteration $\vec{\theta}_{\rm ID}^{\,(i)}$ based on Sec.~\ref{sec:updateforms} and set $\vec{\chi}_{A}^{\,(i)}(t=0) =\vec{\chi}_{A,T}$. 
Unless otherwise indicated, by default we start simulations at apastron (box 2).
\item We numerically evolve the system up to a time $5\pi/\Omega_0$.  In practice this corresponds to roughly $3$ to $5$ orbits, depending on the eccentricity of the orbit (boxes 3-4).
\item We fit the numerical data for the orbital parameters $\vec{\theta}_{\rm orb}^{\,(i)}$ and read off the spins $\tilde{\chi}_{A}^{\,(i)}(t=t_T)$ (box 5).
\item If the orbital parameters $\vec{\theta}_{\rm orb}^{\,(i)}$ are not equal to the target parameters $\vec{\theta}_{\rm orb,\,T}{}$ within some tolerance (box 6), we update the input parameters with Eqs.~\eqref{newOmega0}-\eqref{newD0} and the spins with Eq.~\eqref{rotation_BHNR} (box 7) and start a new simulation (box 3).
\item Once the eccentricity threshold is achieved (Box 6), we evolve the system to merger (box 8).
\end{enumerate}

For simplicity, we adopt the eccentricity threshold imposed by the initial implementation of the parameter control loop in SpEC aiming at vanishing eccentricity~\cite{Buonanno:2010yk}: $|e_T-e^{(i)}|\leq 7\times 10^{-4}$. This tolerance is set just below the eccentricity value for which spin-induced oscillations dominate the eccentricity-induced oscillations in $\dot{\Omega}$. Therefore, it is a sufficient constraint for a parameter control loop that relies on $\dot{\Omega}$.

Even though we only set a tolerance threshold on the eccentricity, in practice we find that all orbital and spin parameters converge to their target value reasonably well.
Adding further tolerance thresholds on spins or further orbital parameters is straightforward. 
Having established the iterative parameter update procedure, we turn to the post-Newtonian equations required for the fits and updates of boxes 4 and 6.


\subsection{Eccentric post-Newtonian dynamics}
\label{sec:pn}

We fit the numerical data with the Keplerian equations of eccentric motion augmented with post-Newtonian corrections as given in Ref.~\cite{pw}. 
We use these post-Newtonian equations to extract the Keplerian parameters $ \vec{\theta}_{\rm orb}$ from numerical data and to compute the inverse operator ${\cal{G}}$ in Eq.~\eqref{eq:deltax_NR}.

A generic Keplerian orbit at Newtonian order, 0PN, is characterized by its semimajor axis $a$, eccentricity $e$, and mean anomaly $\ell$  defined at the epoch $t=0$.  
These three quantities are related via Kepler's equations: $\bar{\Omega} t=u(t)-e\sin{u(t)} - \ell$ and $\bar{\Omega} = \sqrt{M/a^3}$. The eccentric anomaly, $u(t)$, is obtained by inverting Kepler's equation, which in the small-eccentricity limit can be approximated as: $u(t) = \bar{\Omega} t + \ell$.
To next, 1PN order, the Keplerian parameters and all equations obtain corrections. 
These can be expressed in closed form and as a function of the Newtonian orbital parameters $(a,e,\ell)$. 
We present these equations and their derivation in Appendix~\ref{appendixa}.
The pertinent point is that there is a closed-form expressions for the (derivative of the) orbital angular frequency as a function of time, $\dot{\Omega}(t;a,e,\ell).$ 

\subsection{PN trajectory fitting model}
\label{sec:fittingmodel}

We characterize the numerical trajectory through the derivative of the orbital angular velocity $\dot{\Omega} (t)$ as shown in Fig.~\ref{fig:method}. 
This choice follows Ref.~\cite{Buonanno:2010yk}, which states that eccentricity-induced oscillations become much more pronounced, and therefore easier to fit, in $\dot{\Omega}(t)$ than in $\Omega (t)$.

In the post-Newtonian framework, $\dot{\Omega} (t)$ deviates from zero (circular orbit) due to three effects 
\begin{align}
\dot{\Omega}(t) &= \delta\dot{\Omega}_{\RR}(t)+\delta \dot{\Omega}_{\s}(t)+\delta \dot{\Omega}_{e}(t)\label{eq:omegadotPN}\,.
\end{align}
The first term in Eq.~\eqref{eq:omegadotPN} encodes radiation reaction effects which, including 1PN corrections, read
\begin{align}
\delta\dot{\Omega}_{\RR}(t) =& \frac{96}{5}\eta M^{7/2} \left(D_0^4-\frac{256}{5}\eta M^3 t\right)^{-11/8}\label{omegadotRR}\nonumber\\
        &- \frac{48}{5}\eta (3 - \eta) M^{9/2} \left(D_0^4-\frac{256}{5}\eta M^3 t\right)^{-13/8},
\end{align}
where $D_0$ is the initial BH separation. 

The second term in Eq.~\eqref{eq:omegadotPN} encodes 2PN spin-spin interactions~\cite{Buonanno:2010yk}
\begin{align}
  \delta\dot{\Omega}_{\s}(t) &= \frac{1}{2} \left[ \left(\vec{S}_0\cdot\bm{\hat{n}_0}\right)^2 + \left(\vec{S}_0\cdot\bm{\hat{\lambda}_0}\right)^2 \right] \sin (2\bar{\Omega}t + \phi)\,,\label{omegadotSS}
\end{align}
which induce frequency modulations at twice the orbital frequency even for circular orbits. 
We use the definitions of Ref.~\cite{Buonanno:2010yk}, where $\vec{S}_0$ is the mass-weighted spin vector, $\hat{n}_0$ is the unit normal to the orbital plane, and $\hat{\lambda}_0$ is orthogonal to $\hat{n}_0$ and the angular orbital momentum vector, $\hat{L}$. 
Subscripts indicate that these unit vectors are defined at $t=0$.  
The phase offset $\phi$ is a known function of $\vec{S}_{0}$ given by Eq. (50) in Ref.~\cite{Buonanno:2010yk}. 

The final term in Eq.~\eqref{eq:omegadotPN} encodes eccentric effects, therefore it is a function of the orbital parameters. To 1PN order~\cite{pw}
\begin{equation}
\label{omegadote1PN}
    \delta \dot{\Omega}_e(t) = - \frac{A \sin u(t)(1 - \tilde{e} \cos u(t))}{(1 - e_t\cos u(t))^3(1 - e_\phi \cos u(t))^2} \,,
\end{equation}
where $A, e_t, e_\phi, \tilde{e}, u(t)$ are known functions of the Keplerian orbital parameters $(a,e,\ell)$ given in Appendix~\ref{appendixa}. 

Appendix~\ref{Omegae_derivation} also presents a derivation of Eq.~\eqref{omegadote1PN} based on Ref.~\cite{pw}.
Crucially, $\delta \dot{\Omega}_e(t)$ at this order depends only on $(a,e,\ell)$, so we can use this expression to fit numerical data and obtain estimates for the orbital parameters.

Motivated by the three terms of Eq.~\eqref{eq:omegadotPN}, we define the following model to fit the numerical data, shown by the dashed curve in Fig.~\ref{fig:method}
\begin{align} 
    \dot{\Omega}_{\rm model} (t) &= C_1 (T_c - t)^{-11/8} + C_2(T_c-t)^{-13/8} \nonumber \\
    &+ C_3 \cos\alpha(t) + C_4 \sin\alpha(t) \nonumber \nonumber \\
    &- \frac{A \sin u(t)(1 - \tilde{e} \cos u(t))}{(1 - e_t\cos u(t))^3(1 - e_\phi \cos u(t))^2}\,.\label{eqn:fittingmodel}
\end{align}
The first line of Eq.~\eqref{eqn:fittingmodel} captures radiation-reaction. 
It depends on amplitude parameters $C_1$ and $C_2$ that we fit for. 
The coalescence time, $T_c$, is calculated from the evolution data using using a 0PN approximation for the time to merger
\begin{equation}
    T_c = t_T + \frac{5M^2}{256\eta (M\Omega(t_T))^{8/3}} \,,
\end{equation}
where $\Omega(t_T)$ is the angular frequency at time $t=t_T$, read off from the numerical data.

The second line captures spin-spin interaction effects of the functional form of Eq.~\eqref{omegadotSS}. 
It again depends on amplitude parameters $C_3$ and $C_4$ that we fit for, while $\alpha(t)$ is a known function that is directly calculated from the numerical data, see Eq.~(50) in Ref.~\cite{Buonanno:2010yk}.
Finally, the third line of Eq.~\eqref{eqn:fittingmodel} captures eccentricity-induced oscillations, the primary objective of this work.
All quantities appearing in this term are known functions of the orbital parameters $a, e, l$ that we fit for.

Formally, Eq.~\eqref{eqn:fittingmodel} should contain additional PN corrections to each of the three effects beyond the PN order in which they each appear. 
In practice, we find that this expression is adequate for fitting numerical data with the goal of obtaining updated input parameters.

\subsection{Fitting the numerical data}
\label{reference_time}

We fit the numerical data with Eq.~\eqref{eqn:fittingmodel} to extract the orbital parameters, $\vec{\theta}_{\rm orb}=(a,e,\ell)$. This is a seven-parameter nonlinear fit, but the values of four of the parameters $(C_1,C_2, C_3, C_4)$ are not used in any of the subsequent formulas.
First, we must choose a reference time, $t_T$, the beginning of the fitting interval. 
In Fig.~\ref{fig:method}, this interval is indicated by the white region to the right of the dotted line ($t = t_T$). 
Our motivation for choosing $t_T$ is twofold. 
First, we only want to fit the trajectory after junk radiation has ended. 
Second, we require $t_T$ to be a multiple of the orbital period such that the BHs have returned as closely as possible to their starting positions. 
Both requirements are met with $t_T$ set to 1 orbital period, in practice occurring on the order of $\sim 300$\,M into the simulation evolution. Currently, we cap $t_T$ at $500$ M but could update the determination of the end of junk radiation using Ref.~\cite{pretto2024automateddeterminationendtime}. We implement a cap in the event $5\pi / \Omega_0$ yields an inefficiently large $t_T$. 

Next, we must decide how long to evolve the BBH after $t_T$. 
We motivate this choice with three arguments. 
First, the fitting window must be short to mitigate computational cost for the trial simulation whose main goal is to evaluate Eq.~\eqref{eq:deltax_NR} and provide improved initial parameters. 

Second, the eccentricity, semimajor axis, and mean anomaly are PN parameters that evolve due to radiation reaction and become ill defined close to merger. 
The fitting window must therefore both be short and as far away from merger as possible.
Third, since we are fitting harmonic functions, at least a few orbits are required to achieve enough precision. 
In practice, we evolve the trial simulation for a little less than 5 orbits up to a final time $T_{\rm end} = 10\pi/\Omega_0$, 
calculated based on the initial orbital angular frequency. The PN fit then occurs between times $t_T$ and $T_{\rm end}$.

\subsection{Updating the input parameters}
\label{sec:updateforms}

At this stage, we have performed a trial simulation with input parameters $\vec{\theta}_{\rm ID}^{\,(i)}$ and spins $\vec{\chi}_{A}^{\,(i)}(t=0)$. 
Using Eq.~\eqref{eqn:fittingmodel}, we have fitted the evolved trajectory to obtain orbital parameters $\vec{\theta}_{\rm orb}^{\,(i)}$. 
Using the simulation data, we also read off spins $\tilde{\chi}_A^{(i)}(t=t_T)$. 
The next step is to use this information to compute updated initial-data parameters for the next iteration.  
We introduce two independent update procedures for the orbit and spins.  

\subsubsection{Orbit}
\label{orbit-update}

Given the orbital parameters of the current iteration $\vec{\theta}_{\rm orb}^{\,(i)}$, we calculate initial data parameters for the next iteration $\vec{\theta}_{\rm ID}^{\,(i+1)}$. 
First, we need a mapping from $\vec{\theta}_{\rm orb}$ to $\vec{\theta}_{\rm ID}$.  
We evaluate $\Omega(\vec{\theta}_{\rm orb},t)$, $v_r(\vec{\theta}_{\rm orb},t)$, and $D(\vec{\theta}_{\rm orb}, t)$ using Eqs.~(\ref{eqn:D0_formula})-~(\ref{inp_param_eqns-omega0}), which are derived from 1PN equations of motion in Appendix~\ref{input_derivation}. 
Per Eq.~\eqref{eq:deltax_NR}, the difference between these functions evaluated for $\vec{\theta}_{\rm orb}^{\,(i)}$ and $\vec{\theta}_{\rm orb, T}$ at time $t = 0$ is the amount we need to correct the current iteration initial data parameters by, yielding
\begin{align}
    \Omega_0^{\,(i+1)} &= \Omega_0^{\,(i)} + \left[\Omega\left(\vec{\theta}_{\rm orb,T}, 0\right) - \Omega\left(\vec{\theta}_{\rm orb}^{\,(i)}, 0\right)\right]\,,\label{newOmega0}\\
    v_{r,0} ^{\,(i+1)}&= v_{r,0}^{\,(i)} + \left[v_r\left(\vec{\theta}_{\rm orb,T}, 0\right) - v_r\left(\vec{\theta}_{\rm orb}^{\,(i)}, 0\right)\right]\,,\\
    D_0^{\,(i+1)} &= D_0^{\,(i)} + \left[D\left(\vec{\theta}_{\rm orb,T}, 0\right) - D\left(\vec{\theta}_{\rm orb}^{\,(i)}, 0\right)\right] \label{newD0}\,.
\end{align}

\subsubsection{Spin}
\label{spin-update}
\begin{figure}[] \label{fig:spinupdateschematic}
\includegraphics[width=0.75\columnwidth,clip=true]{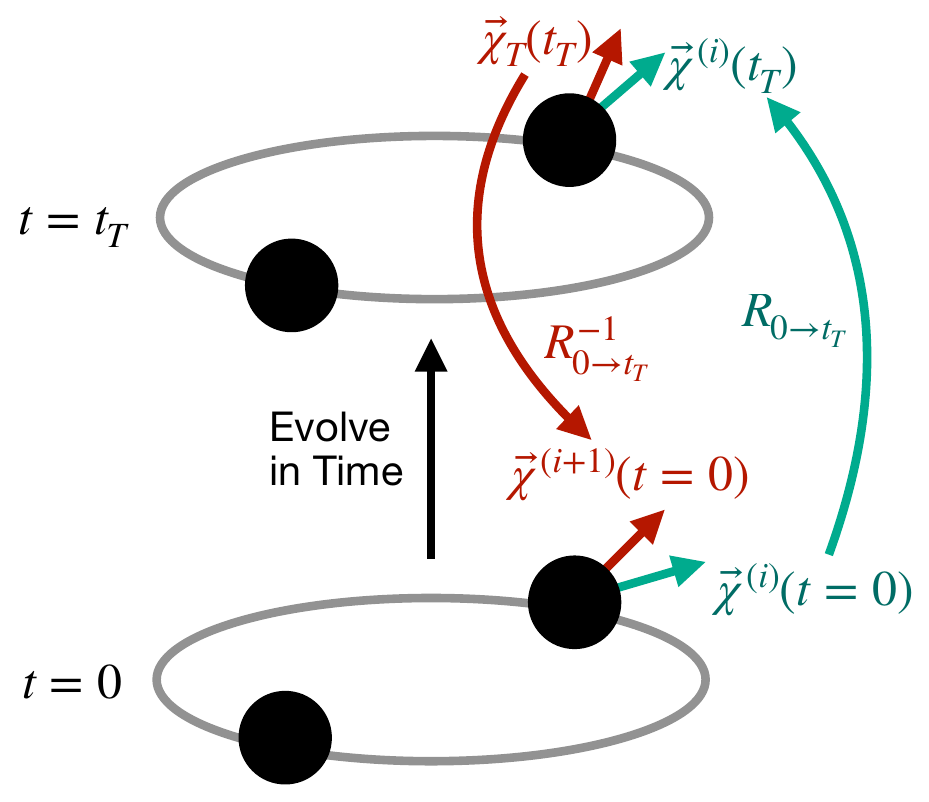}\\
\caption{\label{fig:spinupdateschematic} Schematic for spin vector updates described in Sec.~\ref{spin-update}. The bottom diagram shows the BBH at $t = 0$ and the top diagram shows the BBH at $t = t_T$, one orbit later. 
The vectors in the bottom panel represent the initial spin vector for the current iteration, $\vec{\chi}^{\,(i)}(t = 0)$, and the updated initial spin vector for the next iteration, $\vec{\chi}^{\,(i+1)}(t = 0)$. 
The top panel shows the target spin vector, $\vec{\chi}_{T}(t = t_T)$, and the evolved spin vector at time $t = t_T$, denoted $\vec{\chi}^{\,(i)}(t = t_T)$. The green and red arrows map vectors between times $t = 0$ and $t_T$ via rotation matrix $R_{0\to t_T}$ and its inverse, respectively. The spin arrows and text at times $t=0$ and $t=T$ correspond to the relevant rotation matrix and arrow. Though the figure depicts only spin on
  one BH for clarity, we apply the same procedure to both BHs' spins.
} 
\end{figure}

\begin{figure*}[]
\includegraphics[width=0.85\textwidth,clip=true]{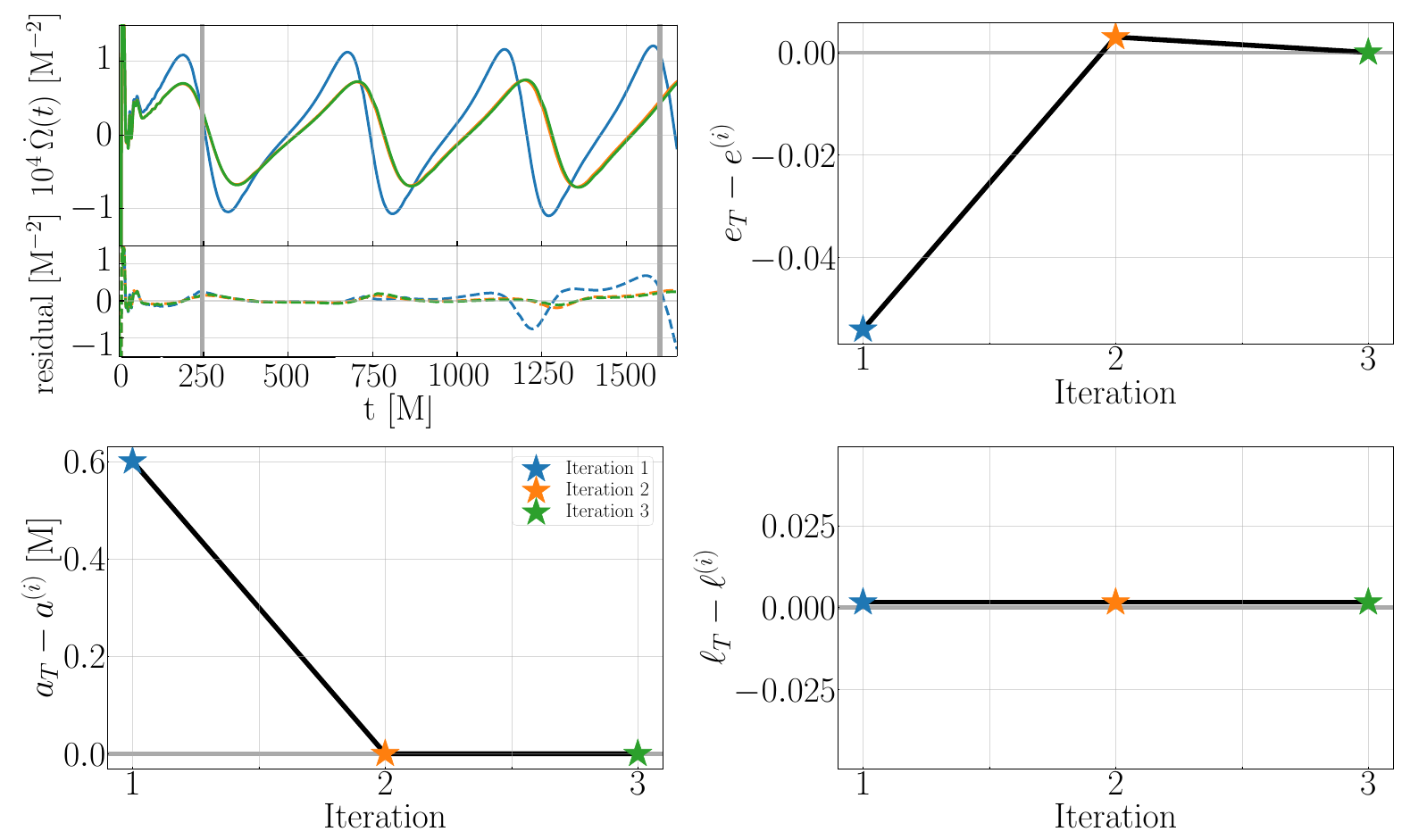}\\
\caption{ \label{fig:exampleplots} Application of the parameter control loop outlined in Fig.~\ref{fig:eccflowchart} for an equal-mass binary with vanishing spins and target orbital parameters $\vec{\theta}_{\rm orb}{_{,T}} = \{a_T= 15\,{\rm M}, e_T=0.2, \ell_T=\pi\}$. 
The top left panel shows the NR data for each iteration of the control loop (top subpanel) and the residuals between the PN fit and NR data for each iteration (bottom subpanel).
The rightmost gray vertical line indicates $t = t_T$. The PN fitting window occurs between the two vertical gray lines.
The difference between the achieved, i.e., fitted from the NR data, orbital parameters in each iteration $i$ and the target parameters is plotted in the remaining panels: eccentricity (top right), semimajor axis (bottom left), and mean anomaly (bottom right).
All parameters converge to their target value after $3$ iterations
}
\end{figure*}

The spin update procedure is applied to both BHs' spins; for clarity we discuss only one case and drop the BH subscripts.
We recall that the target spins $\vec{\chi}_{T}$ are given in the coorbiting BH frame at $t=t_T$, whereas the numerical evolution yields spin vectors in the inertial frame.
Thus, the first step is to rotate the target spins into the inertial frame.
We define this rotation matrix as $R_{\rm BH\to in}(t=t_T)$, constructed using the inertial frame basis of $\hat{n}, \hat{\lambda},$ and $\hat{L}$
introduced in the context of Eq.~(\ref{omegadotSS}) but defined at time $t_T$. 
Next, we correct for the change in the spin directions during the evolution from $t = 0$ to $t = t_T$. 
We introduce another rotation matrix, denoted $R_{0\to t_T}$, shown in Fig.~\ref{fig:spinupdateschematic}, as the rotation from $\vec{\chi}(t = 0)$ to $\vec{\chi}(t = t_T)$.
We then apply the inverse rotation $R_{0\to t_T}^{-1}$ to $\vec{\chi}_T$ to approximate the position of that vector at the beginning of the simulation, $t = 0$.  
Combining the two rotations yields the initial spin of the next iteration
\begin{equation} \label{rotation_BHNR}
    \tilde{\chi}^{\,(i+1)} (t=0) = R_{0\rightarrow t_T}^{-1} \,R_{\rm BH \rightarrow in}(t=t_T)\,\vec{\chi}_T(t=t_T)\,.
\end{equation}
The derivation and explicit forms of rotation matrices $R_{0\to t_T}^{-1}$ and $R_{\rm BH\to in}(t=t_T)$ are given in Appendix \ref{apx:spin_update_derivation}.

\section{Eccentric, nonprecessing orbits}
\label{sec:aligned}

We begin with an example of the parameter control process. Figure~\ref{fig:exampleplots} shows results for an equal-mass binary, target orbital parameters $\vec{\theta}_{\rm orb}{_{,T}} = \{a_T= 15\,{\rm M}, e_T=0.2, \ell_T=\pi\}$, and vanishing spins.\footnote{Unless otherwise indicated, we set
    $\ell_T$ to apastron as this choice leads to more stable evolutions. }
The top left panel shows the numerical data for $\dot{\Omega}(t)$ for each of the three iterations required to achieve the eccentricity threshold. 
Each iteration follows the control process shown in Fig.~\ref{fig:eccflowchart} to produce fitted orbital parameters for the numerical data and update the initial data parameters of the next iteration accordingly.

\begin{figure*}[]
\includegraphics[width=0.95\textwidth,clip=true]{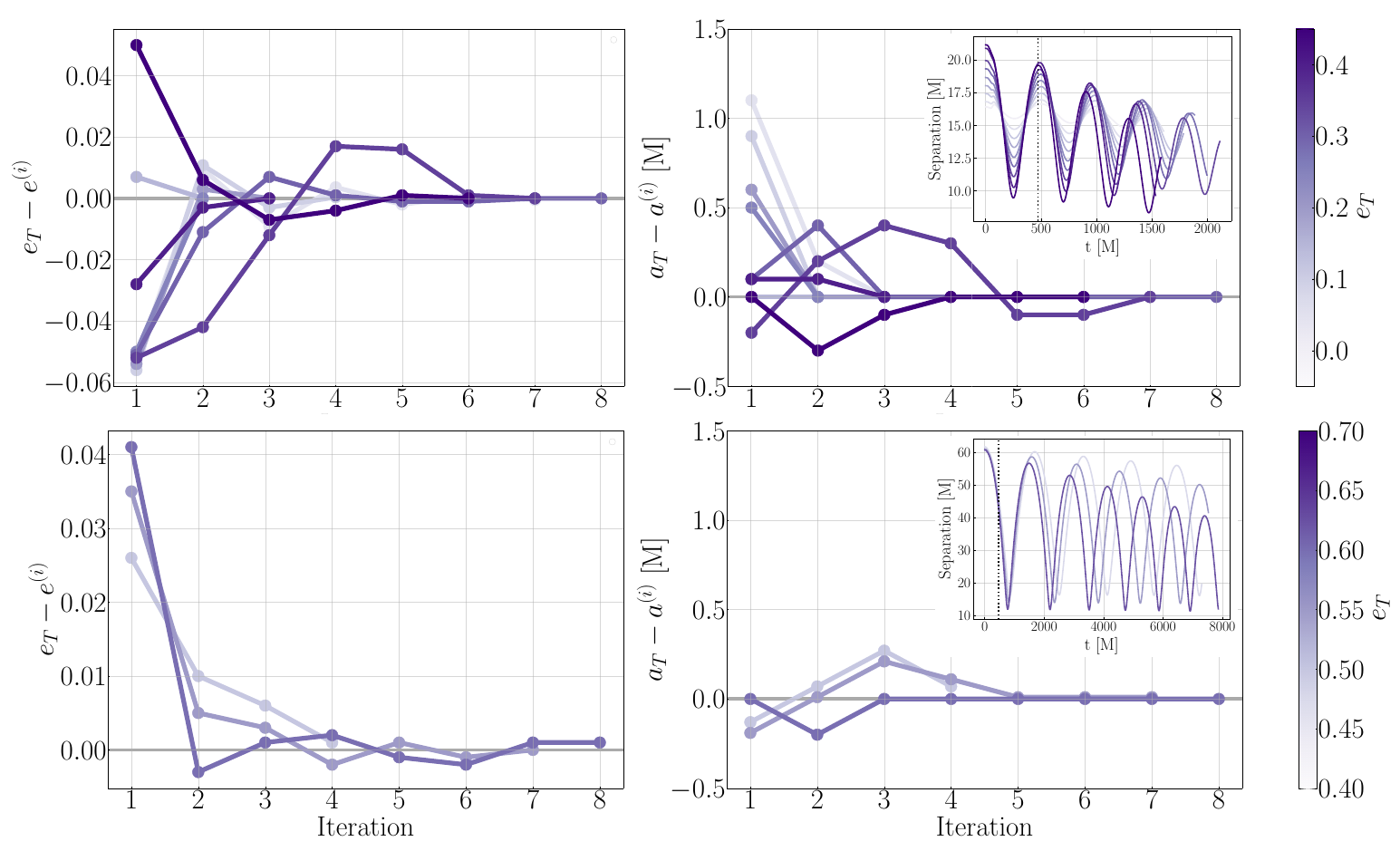}\\
\caption{ \label{fig:allecc}Impact of target eccentricity value on parameter convergence. 
We plot the difference between the fitted and target eccentricity (left panel) and semimajor axis (right panel) at each iteration for binaries with equal masses, vanishing spins, $a_T = 15$ M (top) or $r_{a,T}=60$ M (bottom), and different values of target eccentricity $e_T$ (color bar).
We achieve parameter convergence for all the cases we attempted, including eccentricities up to $0.65$. 
For $e_T>0.65$, the BHs risk head-on collision, which terminates the control loop prematurely before obtaining $\vec{\theta}_{ID,T}$. 
Thus we omit cases $e_T>0.65$ in this work.
As $e_T$ increases to 0.65, more iterations are required for parameter convergence. 
We also inlay separation over time plots based on the final (most converged) iteration of the control loop.
}
\end{figure*}

The remaining panels show the difference between the achieved (fitted from the NR data) and the target eccentricity (top right), semimajor axis (bottom left), and mean anomaly (bottom right) as a function of iteration. Here and in subsequent figures, $a^{(i)}, e^{(i)}$ and $\ell^{(i)}$ refer to the fitted orbital parameters in each iteration $i$ of the control loop. 
Even though we only impose a tolerance threshold on the eccentricity, all orbital parameters converge to their target values after $3$ iterations.
In fact, the mean anomaly, $\ell$, converges almost immediately. 
Since this behavior is typical in subsequent explorations, we omit the mean anomaly from now on.
Moreover, the semimajor axis converges in 1 fewer iteration than the eccentricity.
This justifies continuing to consider only the eccentricity in the parameter control tolerance.

Convergence in all parameters is monotonic, as in every iteration they are closer in absolute value to their target value.
However, while the eccentricity is larger than the target value in the first iteration, it is slightly smaller in the second iteration. 
This suggests that the updated parameters from the first to the second iteration ``overshoot" slightly.
This characteristic ``oscillatory" convergence is typical for the eccentricity parameter.

\subsection{Impact of target eccentricity value}

Having demonstrated the parameter control loop in practice for one binary, we now explore the impact of the target eccentricity value on convergence. 
We again use equal masses, vanishing spins, and vary $e_T$ between $0$ and $0.65$. 
In highly eccentric binaries, the periastron passage might bring the BHs close enough that the post-Newtonian approximation becomes inaccurate or the system directly merges. 
The BH separation at periastron depends on whether we increase the eccentricity at a fixed semimajor axis $a$ or apastron separation $r_a = a(1+e)$. 
For binaries with $e_T < 0.5$, we find that the former suffices and fix $a_T = 15$ M. 
For binaries with $e_T \geq 0.5$, we instead increase the eccentricity at a fixed $r_{a,T} = 60$ M. 

Figure~\ref{fig:allecc} shows parameter convergence at fixed $a_T$ and $e_T < 0.5$ (top) and fixed $r_{a,T}$ and $e_T \geq 0.5$ (bottom). 
We show the difference between the target and fitted eccentricity (left panels) and semimajor axis (right panels) as a function of iteration and colored by the target eccentricity value. 
The inlaid plots show the BH coordinate
separation over time, using the same color scheme. 
In the top row, the separation reflects a constant $a_T$ for each $e_T < 0.5$. 
As the eccentricity increases (darker purple), the average separation stays constant but the apastron and periastron passages become more extreme.  
In the bottom row, all trajectories begin at the same apastron and result in roughly similar periastron values.

In all cases, i.e., for target eccentricty up to $e_T=0.65$, the eccentricity and semimajor axis converge to their target value, with higher eccentricities requiring on average more iterations. 
This is likely because higher eccentricity orbits are more relativistic, thus the post-Newtonian approximation to the dynamics is less accurate. 
Such less accurate fitting equations lead to less efficient initial data parameter updates in each control loop iteration, thus requiring more iterations to converge.

\subsection{Aligned spins}

\begin{figure*}[]
\includegraphics[width=0.9\textwidth,clip=true]{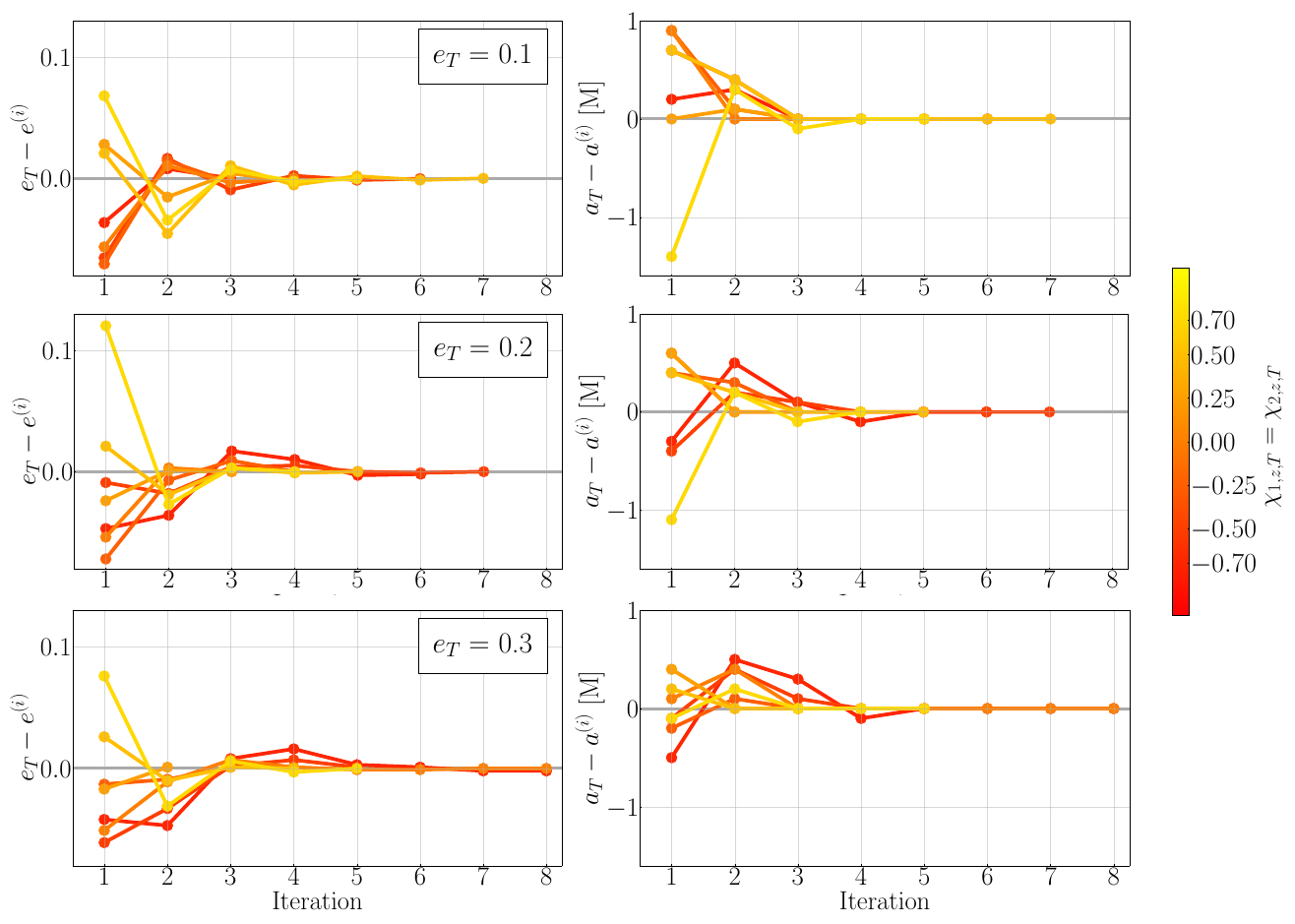}\\
\caption{ \label{fig:alignedspin} Impact of aligned spins on parameter control. We plot the difference between the fitted and target eccentricity (left panel) and semimajor axis (right panel) at each iteration for binaries with equal masses, $a_T = 15$ M, $e_T=\{0.1,0.2,0.3\}$ (top to bottom) and different values of the aligned spin (color bar). We successfully converge to the target parameters for eccentric, aligned spin binaries, with higher eccentricities again requiring more iterations.  
}
\end{figure*}

\begin{figure*}[]
\includegraphics[width=\textwidth,clip=true]{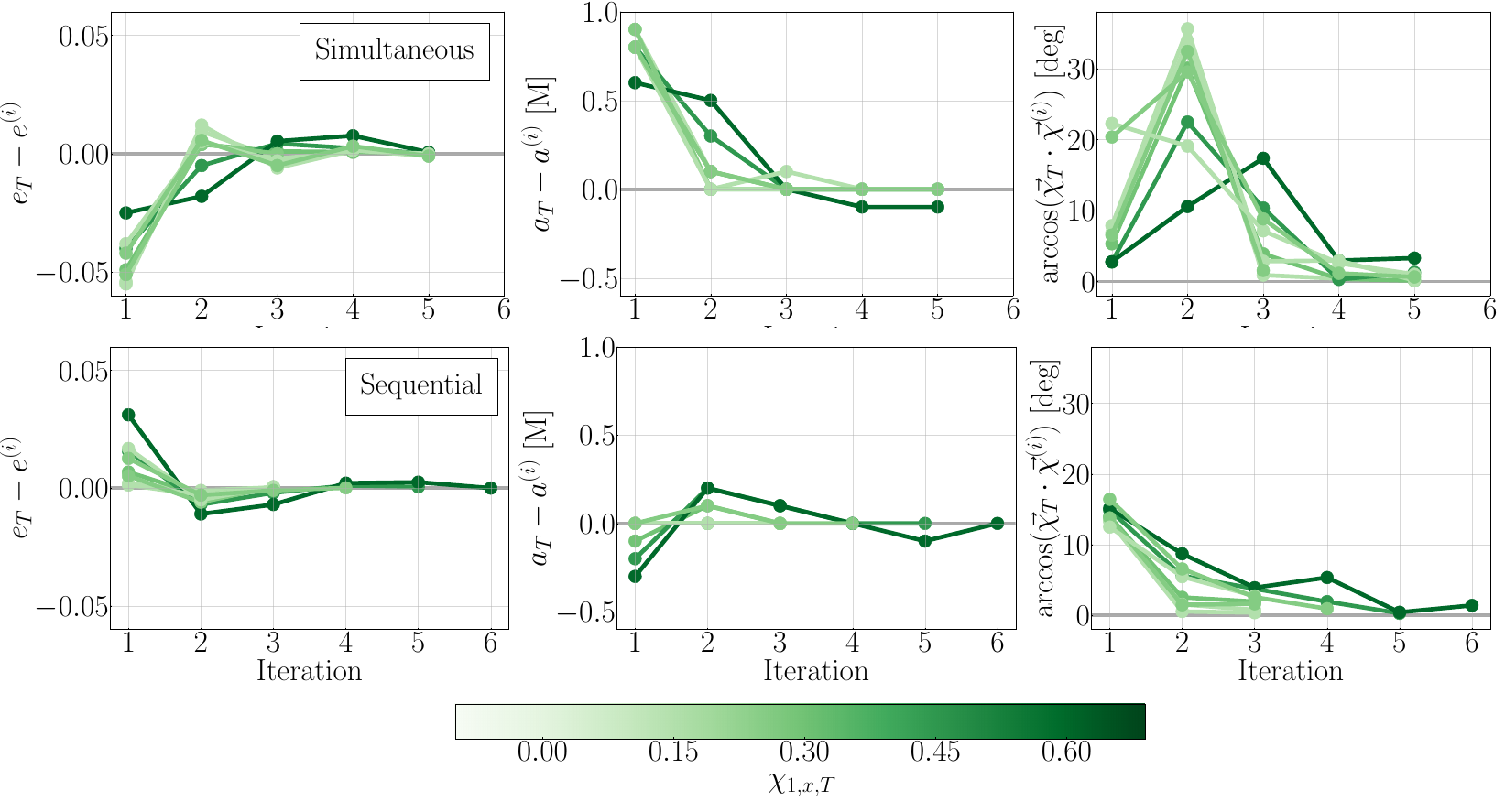}\\
\caption{ \label{fig:methods1_2} Comparing the simultaneous (top) and the sequential (bottom) parameter control approaches. 
We consider equal-mass binaries with $a_T=15$ M, $e_T=0.1$, and different values of $\chi_{1,x,T}$ (color bar).
We show the difference between the target and the achieved value for the eccentricity (left), semimajor axis (middle), and in-plane spin angle, $\arccos(\vec{\chi}_T \cdot \vec{\chi}^{(i)})$ (right).
In the sequential case, we plot results from the second parameter control loop, i.e., after obtaining a simulation with the target orbital parameters and no in-plane spins.
Both methods achieve convergence, though as expected the simultaneous method needs more iterations than the second parameter control loop for the simultaneous method.  
Moreover, the orbital parameters are closer to their target value in the first iteration for the sequential method (bottom) and the spin direction converges monotonically (right).
}
\end{figure*}

Next, we explore the impact of aligned spins. 
Aligned spins remain constant in both direction and magnitude (modulo horizon absorption effects~\cite{Scheel:2014ina}). Therefore the rotations of Fig.~\ref{fig:spinupdateschematic} reduce to the identity, and the spins are not updated during the eccentricity-control iterations. 
However, the spins still affect the binary dynamics and rate of orbital decay and could therefore impact the orbital parameter convergence.
We study binaries with equal masses, $a_T = 15$ M, $e_T = \{0.1,0.2,0.3\}$, and vary the amount of aligned spin $\chi_{z, A, T} = \chi_{z,B, T}$ in $[-0.7, 0.7]$.
Results are shown in Fig.~\ref{fig:alignedspin} for the different target eccentricities (top to bottom) and aligned spin values (color bar).
In all cases, the eccentricity and semimajor axis converge to their target value after $\leq 7 $ iterations.

The initial (iteration 1) evolution consistently under(over)-shoots the target eccentricity for (anti)aligned spins.
Similar to the effect for larger target eccentricities, larger absolute value aligned spin magnitudes require more iterations for convergence.

\section{Eccentric, precessing orbits}
\label{sec:precessing}

In this section, we generalize to precessing binaries.
Going forward, we assume one spinning BH and no aligned component for simplicity.
In-plane spins precess in the inertial frame, causing the BH frame to change orientation. The parameter control loop of Fig.~\ref{fig:eccflowchart} and the updating formulas in Eqs.~\eqref{newOmega0}–\eqref{newD0} and~\eqref{rotation_BHNR} treat the orbital parameters and spins independently for convenience. 
Since the orbital parameters and the spins are updated independently, we begin by exploring two parameter control methods:
\begin{enumerate}
    \item \emph{Simultaneous.} We follow the process of Fig.~\ref{fig:eccflowchart} where in the first iteration we simulate an eccentric, precessing binary and then iteratively update the orbital parameters and the spins, per Eqs.~\eqref{newOmega0}–\eqref{newD0} and~\eqref{rotation_BHNR}.
    \item \emph{Sequential.} We execute the parameter control loop twice.   First, we iterate while updating only the orbital parameters; while doing so, we keep the spins constant at only the aligned-spin component of the target spins.  Subsequently, once the target orbital parameters have been achieved, we restart the loop and update spins and orbital parameters on each iteration.
\end{enumerate}
Figure~\ref{fig:methods1_2} compares parameter convergence for the simultaneous (top) and the sequential (bottom) methods for an equal-mass binary, $a_T=15$ M, $e_T=0.1$ and different values of $\chi_{1,x,T}$, and with $\chi_{1,y,T}=\chi_{1,z,T}=|\vec{\chi}_{2,T}|=0$.
For the sequential case, we present only the second parameter control loop, i.e., the one that updates both the orbital parameters and the spins.
Beyond the eccentricity (left) and the semimajor axis (middle), we show the angle between the target and the actual spin at $t_T$ (right).
As before, all parameter control loops use a tolerance threshold based solely on eccentricity.

Both methods converge to the target parameters, with the simultaneous method needing consistently 4--5 iterations, while the sequential one needs 3--6.
In the sequential case, this estimate does not include the first control loop that sets the in-plane spins to zero and updates only the orbital parameters. From Fig. \ref{fig:alignedspin}, we conclude this first control loop requires anywhere from 3--7 iterations. We recall that greater eccentricities require more iterations to converge to target parameters. 
So overall, in terms of total runtime, we expect the sequential method to be more expensive.
However, there may exist additional applications for this method. For example, we could add in-plane spins in pre-existing simulations such as those in an NR catalog~\cite{Boyle:2019kee}.
Based on this evaluation of the computational efficiency, results in the subsequent subsections are obtained with the simultaneous method.

\begin{figure*}[]
\includegraphics[width=\textwidth,clip=true]{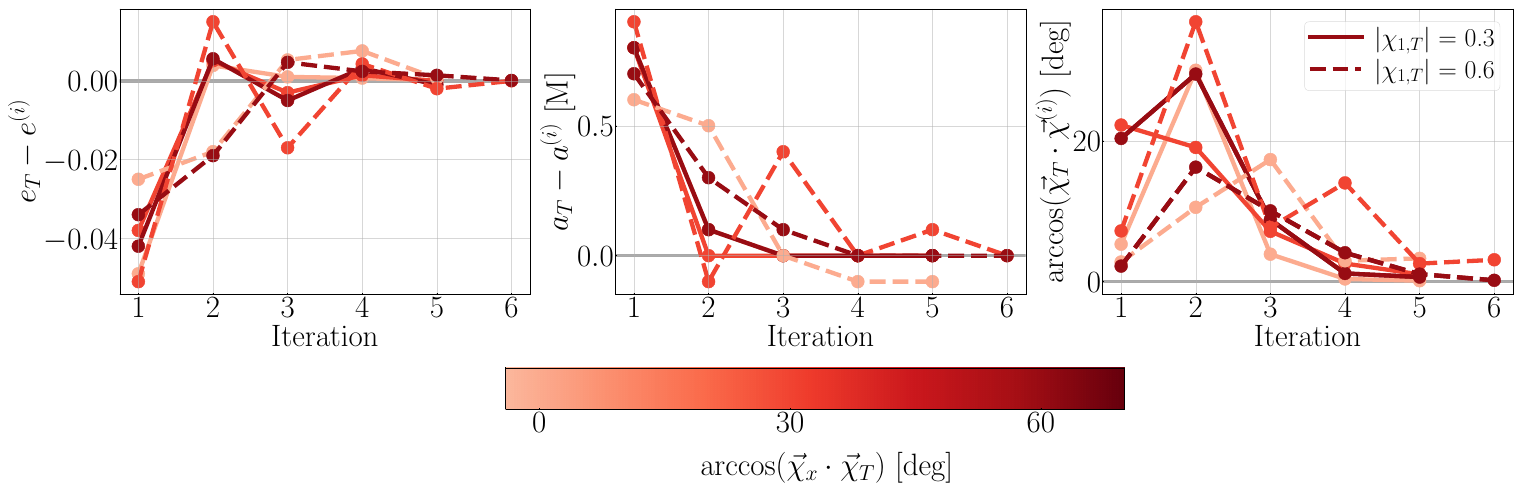}\\
\caption{ \label{fig:xyspins} Parameter convergence for precessing spin and varying spin magnitudes. 
We plot the difference between the target and the fitted eccentricity (left), semimajor axis (middle), and spin direction (right) for different values of the magnitude and direction of the in-plane spin. 
We choose $|\vec{\chi}_{1,T}| = 0.3$ (solid) and $|\vec{\chi}_{1,T}| = 0.6$ (dashed). 
The angle between the target in-plane spin and the $x$-axis is denoted as $\arccos(\vec{\chi}_x \cdot \vec{\chi}_T)$  (color bar). 
In all cases, the parameters converge after $\leq 6$ iterations with behavior depending more on the in-plane spin magnitude than the in-plane spin angle, c.f. also the top panel of Fig.~\ref{fig:methods1_2}, which is similar to this plot except it compares different in-plane spin magnitudes with the same in-plane spin angle.
}
\end{figure*}

Comparing the rate of parameter convergence between the two methods, the sequential method results in orbital parameters closer to their target values at the first iteration.
This is evident for both the eccentricity and semimajor axis.
Even though the second control loop in the sequential method plotted here starts from preconverged orbital parameters, they still differ from their target value.
This is because the first loop achieved the target parameters under zero in-plane spins, which are added in the second loop.
Therefore the difference between the actual and the target eccentricity and semimajor axis in the first iteration is a consequence of the impact of in-plane spins on orbital dynamics.

Simultaneously adjusting the orbital parameters and the spin directions (top) results in spins that are no longer monotonically converging to their target value.
Between the first and the second iteration, and even though both the eccentricity and the semimajor axis are closer to their target values, the spin angle difference increases from $\sim 5-10\,$deg to $\sim20-30\,$deg. 
This is again a consequence of the coupling of orbital and spin dynamics. 
The spin is updated per Eq.~\eqref{rotation_BHNR} for the second iteration assuming the orbit of the first iteration.
Despite this, in the third iteration where the orbit is not substantially updated, the spin angle difference quickly decreases. 
Overall, we achieve differences between the target and the actual spin angle ${\cal{O}}(\rm{deg})$, even though spins are not considered in the tolerance threshold.

\subsection{Effect of in-plane spin}

Let us investigate performance for various
  directions of the spins.  We choose a spin on the primary that is
  tangential to the orbital plane, spanning an angle $\delta$ with the
  x-axis: $\vec\chi_{1,T}=(\chi_1 \cos\delta,\chi_1\sin\delta, 0)$.  We set
  the second BH's spin to zero, $\vec\chi_{2,T}=0$.
  Figure~\ref{fig:methods1_2} investigated performance of our methods
  for $\delta=0$, so let us now discuss different angles $\delta$.

The effect of the in-plane spin magnitude and direction on parameter
convergence are explored in Fig.~\ref{fig:methods1_2} and
Fig.~\ref{fig:xyspins} respectively.  We refer back to the top row of
Fig.~\ref{fig:methods1_2} for varied magnitude for in-plane spin.
Figure~\ref{fig:xyspins} shows the parameter convergence for varied
in-plane spin angle.  We select two in-plane spin magnitudes,
$|\vec{\chi}_{1,T}| = 0.3$ (solid) and $|\vec{\chi}_{1,T}| = 0.6$
(dashed), and vary the angle between $\chi_{1,T}$ and the $x$-axis.
As before we set $\chi_{1,z,T}=|\vec{\chi}_{2,T}|=0$.
Figure~\ref{fig:xyspins} shows that binaries with higher spin magnitudes require more iterations for convergence. 
On the other hand, the impact of the in-plane spin angle is minimal. 
Thus we conclude the number of iterations required to obtain $\vec{\theta}_{\rm orb,T}$ is more dependent on the in-plane spin magnitude than the in-plane spin direction. 
This is also consistent with the top panels of Fig.~\ref{fig:methods1_2}. 

\subsection{Effect of mass ratio}

\begin{figure*}[]
\includegraphics[width=\textwidth,clip=true]{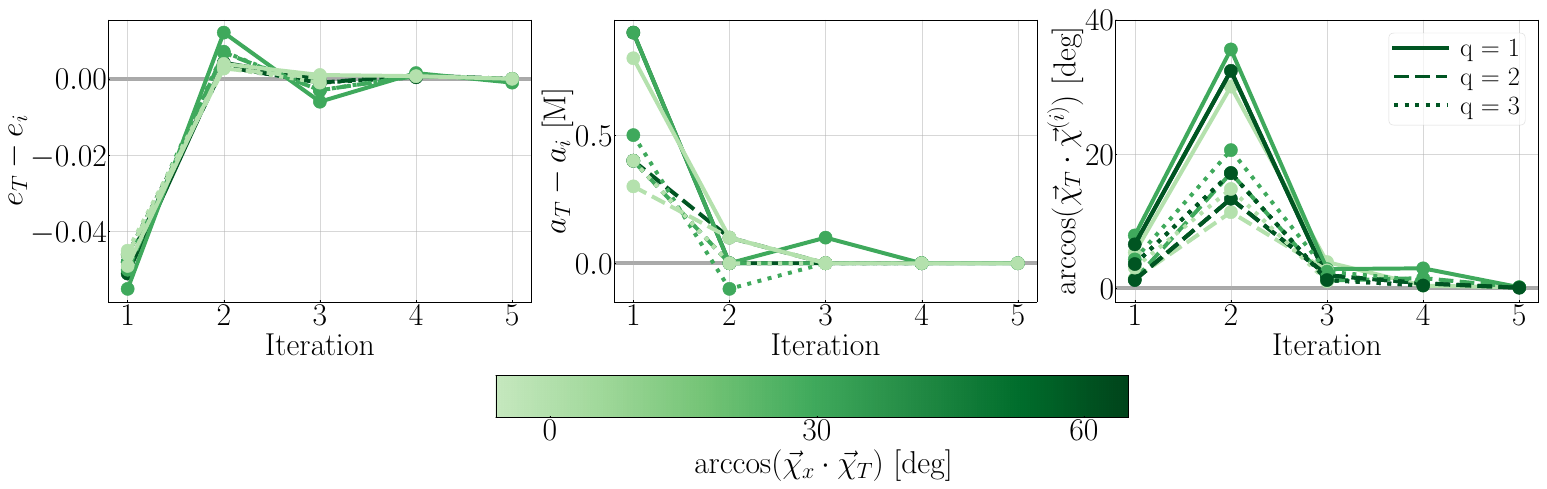}\\
\caption{ \label{fig:q123_precess} Effect of mass ratio on parameter convergence.
We plot the difference between the target and the fitted eccentricity (left), semimajor axis (middle), and spin angle (right) for different values of the mass ratio (line styles) and direction of the target in-plane spin, $\arccos(\vec{\chi}_x \cdot \vec{\chi}_T)$ (color bar). We choose mass ratios $q = 1$, (solid), $q = 2$, (dashed), and $q = 3$, (dotted) and constant target spin magnitude $|\vec{\chi}_{1,T}| = 0.3$. In all cases, the parameters converge after $\leq5$ iterations.
}
\end{figure*}

Finally, we confirm that the above results are robust when changing the mass-ratio.
We revisit the results of Figs.~\ref{fig:methods1_2} and~\ref{fig:xyspins}  for binaries with $a_T=15\,$M, $e_T=0.1$, and varying direction of the in-plane spin with magnitude $|\vec{\chi}_{1,T} | = 0.3$ and extend them to mass ratios $q=2$ and $q=3$. Here $\chi_{1,z,T}=|\vec{\chi}_{2,T}|=0$.
Results are shown in Fig.~\ref{fig:q123_precess} where we obtain convergence in $\leq5$ iterations. 
The mass ratio has a minimal effect on the rate of convergence, something we have also confirmed for zero-spin and spin-aligned configurations.

\section{Discussion}
\label{sec:conclusions}

We have devised a method to achieve numerical BBH simulations in SpEC with target orbital parameters and BH spins beginning at a reference time, $t=t_T$.
Our method is an extension of previous efforts to target vanishing eccentricity. It is based on iteratively performing short numerical simulations, fitting the orbit, and calculating updated initial data parameters until a tolerance threshold is achieved.
We have confirmed convergence for the orbital eccentricity and the spin directions to tolerances of $\mathcal{O}(10^{-3})$ and $\mathcal{O}({\rm deg})$ respectively. 
We have also shown that this approach is effective for binaries with mass ratios $q \leq 3$, eccentricities $e \leq 0.65$, and spin magnitudes $\chi \leq 0.75$ in $\leq7$ iterations.

Overall, convergence is achieved with fewer iterations for smaller spin magnitudes and eccentricities, while the mass ratio has minimal impact on the rate of convergence.\footnote{Each iteration's computational time and resources depend on the target orbital parameters. For moderate spin magnitudes and comparable masses at moderate eccentricity, one iteration takes only a few hours.  For large eccentricity (where the orbital period is longer), or for large spin magnitudes and/or mass-ratio (where the code runs more slowly), the time per iteration can reach several days.  Furthermore, the first few iterations of the parameter control are run at a coarser resolution. When the eccentricity is close to the target, SpEC switches to a higher resolution. This reduces computational cost for the first few iterations of the parameter control. 
}
The number of iterations required sensitively depends on the initial parameters of the first iteration.
In the absence of a better solution, we calculate them based on the target orbital parameters and the 1PN equations of motion.
A more robust procedure to calculate the initial data parameters of the first iteration would improve convergence.
For example, rough initial data parameters for quasicircular simulations can be obtained for SpEC based on a Gaussian process regression fitted on the existing extensive catalog of quasicircular simulations~\cite{Boyle:2019kee}.
Once an appropriately large eccentric catalog is available, similar techniques could be adopted for calculating input parameters of future eccentric simulations.
Other approaches include using higher-PN order or effective-one-body equations of motion~\cite{Nagar_2021}. 

Currently and inspired by the quasicircular case, the tolerance threshold for terminating the parameter control loop is an accuracy of $7\times10^{-4}$ in eccentricity while the other parameters are ignored.
In practice, this choice still achieves good convergence for all parameters, notably the semimajor axis and spin directions.
In general, the appropriate tolerance threshold depends on the application.
If future applications, such as building surrogate models that require simulations in sparsely covered regions of the parameter space~\cite{Varma_2019}, require greater precision in other parameters, adjusting the tolerance threshold in the parameter control loop is trivial.

Finally, a caveat in studying eccentric orbits in general relativity is that no unique, gauge-invariant definition of eccentricity exists~\cite{Shaikh_2023}.
For our purposes, we define the eccentricity of the orbit as simply the Keplerian parameter that we expand upon for higher orders of PN in our equations of motion. 
Since we work to 1PN, we can also self-consistently calculate different orbital eccentricity parameters, e.g., Eq.~\eqref{eq:def-et}. 
One constraint of our method is that we define the eccentricity based on orbital data, rather than the emitted gravitational wave~\cite{Buonanno:2010yk}.
Extracting the gravitational wave during the parameter control loop is possible. Still, it requires either evolving the simulation for longer in each iteration or using techniques such as Cauchy characteristic extraction~\cite{Bishop:1996gt, Bishop:1997ik, Handmer:2014qha, Barkett:2019uae, Moxon:2020gha, Moxon:2021gbv}.
We leave the exploration of ways to define the eccentricity from the gravitational wave signal to future work.

We have shown that we can produce vacuum numerical simulations of BBHs in SpEC with target orbital and spin parameters. 
Such precisely tuned simulations could be used to compare waveforms produced by different numerical relativity algorithms~\cite{Lovelace:2016uwp}, fill in sparsely covered regions of the parameter space for constructing faithful surrogate models to NR, or more broadly the study the dynamics of eccentric, precessing systems in full general relativity.

\acknowledgments
The computations presented here were conducted partly in the Resnick High Performance Computing Center, a facility supported by Resnick Sustainability Institute at the California Institute of Technology. 
Computations for this work were also partly performed with the Wheeler cluster at Caltech. We thank Saul Teukolsky for discussions on fitting methods and the SXS group for advice on using SpEC.
This work was supported by the Sherman Fairchild Foundation, and by NSF
Grants No. PHY-2309211, No. PHY-2309231, and No. OAC-2209656 at Caltech.
\section*{Data Availability}
The data that support the findings of this article are not publicly available because they are owned by a third party and the terms of use prevent public distribution. The data are available from the authors upon reasonable request. 

\appendix

\section{POST-NEWTONIAN FORMULAS FOR A GENERIC BINARY ORBIT}

\label{appendixa}

In this Appendix we present various post-Newtonian formulas used in the main text.

\subsection{Eccentric post-Newtonian dynamics}
\label{sec:pn-app}

In NR, equations of motion at Newtonian order are not sufficient to fully describe BBH evolutions. Therefore, we must introduce corrections to the Keplerian equations of eccentric motion.

We characterize a generic Keplerian orbit at Newtonian order (0PN) by its semimajor axis $a$, eccentricity $e$, and mean anomaly $\ell$. We relate these three quantities through Kepler's equations
\begin{equation} \label{eq:u_nonlin}
    \bar{\Omega} t=u(t)-e\sin{u(t)} - \ell
\end{equation}and $\bar{\Omega} = \sqrt{M/a^3}$. We define $\ell$ as the mean anomaly at the epoch $t=0$.

An approximation of this expression for small eccentricity is $u(t) = \bar{\Omega} t + \ell$.
Furthermore, we define the separation of the two bodies in orbit as: $r(t) = a(1 - e\cos u(t))$.

Next, we introduce 1PN corrections to the semimajor axis, $a$, and eccentricity, $e$. We denote 1PN-corrected quantities with a bar ($\bar{a}$, $\bar{e}$, etc.). We relate the 1PN-corrected quantities to Newtonian quantities via 
\begin{align}
    a &= \bar{a} \left[ 1 - \frac{(8 - 3\eta)M}{2\bar{a}} \right]\,,\\
    e &= \bar{e} \left[ 1 + \frac{(8 - 3\eta)M}{2\bar{a}} \right]\,,\\
    r &= \bar{r} - \frac{1}{2}(8 - 3\eta)M\,.
\end{align}
These preserve Newtonian relations in 1PN: $\bar{r}(t) = \bar{a}(1 - \bar{e}\cos u(t))$ and $\bar{\Omega}\equiv\sqrt{M/a^3}$. Because we are interested in a local time fit (covering only about three orbits), whereas $a$, $e$, $\bar\Omega$ and $\bar r$ change on the
radiation-reaction time scale, we assume these quantities to be constant.  
We approximate the 1PN solution to Kepler's equation as 
\begin{equation}
    u(t) = \sqrt{\frac{M}{a^{3}}} \left[1 -\frac{(9-\eta)}{2} \frac{M}{a} \right] t + \ell \,,\label{eq:Kepler1pnLowEcc}
\end{equation}
where the 1PN correction depends on the symmetric mass ratio $\eta$. Here, we have assumed a small eccentricity in Eq.~\eqref{eq:u_nonlin} such that $e\sin u(t)$ goes to zero. Then we insert the 1PN expression for $\bar{\Omega}$. Although numerically inverting Eq.~(\ref{eq:u_nonlin}) is more accurate for larger eccentricities, it is more computationally expensive than utilizing Eq.~(\ref{eq:Kepler1pnLowEcc}). 

In post-Newtonian theory, noncircularity is described with multiple eccentricity parameters: the radial $\tilde{e}$, temporal $e_t$, and angular eccentricity $e_\phi$. We can similarly define another set of 1PN quantities in terms of the radial eccentricity parameter, $\tilde{e}$. This second set of 1PN quantities are denoted by a tilde ($\tilde{a}$, $\tilde{e}$, etc.). We can relate this set of 1PN-corrected quantities to the Keplerian orbital parameters via
\begin{align}
    \tilde{a} &= a \left[ 1 + (2-\eta)\frac{M}{a} \right]\,,\label{eq:PNa}\\
    \tilde{e} &= e \left[ 1 - (2-\eta)\frac{M}{a} \right]\,,\label{eq:PNe}\\
    \tilde{r} &= r + (2 - \eta)M\,.
\end{align}
Again, we preserve the separation relationship: $\tilde{r}(t) = \tilde{a}(1 - \tilde{e}\cos u(t))$.
The remaining two PN eccentricity parameters are defined in terms of the Newtonian-order eccentricity 
\begin{align}
    e_t &= e\left[ 1 - \frac{1}{2} (8 - 3\eta) \frac{M}{a}\right]\,,\label{eq:def-et} \\
    e_\phi &= e\left[ 1 + \frac{\eta}{2}\frac{M}{a} \right]\,. 
\end{align}
Defining 1PN orbital parameters allows us to derive PN-corrected trajectory equations for eccentric binary orbits.

\subsection{Deriving Eq.~\eqref{omegadote1PN} for $\dot{\Omega}_e$} 
\label{Omegae_derivation}

The derivation of $\dot{\Omega}_e$ follows Poisson and Will~\cite{pw}. We recall that $\eta$ is the symmetric mass ratio, $a$ is the Keplerian semimajor axis, and $e$ is the Keplerian eccentricity. 
We then introduce the 1PN-corrected orbital quantities to supplement those from Sec.~\ref{sec:pn-app}:
\begin{align}
    P &= 2\pi \sqrt{\frac{a^3}{M}}\left[1 +\frac{(9-\eta)}{2} \frac{M}{a} \right]\,,\label{eq:PNP}\\
    t &= \frac{P}{2\pi}[u (t) - e_t \cos u (t) - \ell] \label{eq:dudt_source}\\
    \tilde{h} &= \sqrt{Ma(1-e^2)} \left[ 1 + \frac{3(1-\eta) + (1+2\eta)e^2}{2(1-e^2)} \frac{M}{a} \right]\label{eq:PNh}\,,
\end{align}
where $P$ and $\tilde{h}$ correspond to the period and angular momentum.
Defining $\phi$ as the orbital angle and recalling that $u(t)$ is the eccentric anomaly, the derivative of the angular velocity to 1PN order is
\begin{align}
    \delta \dot{\Omega}_e &= \frac{d^2\phi}{dt^2} = \frac{d}{dt} \left(\frac{d\phi}{du} \frac{du}{dt}\right)\nonumber \\
        &= \frac{d}{dt} \left[ \frac{P}{2\pi} \frac{\tilde{h}}{\tilde{a}^2} \frac{(1 - e_t \cos u(t))}{(1 - \tilde{e}\cos u(t))^2} \frac{2\pi}{P (1 - e_t\cos u(t))} \right]\nonumber\\
        &= \frac{du}{dt} \frac{d}{du} \left[ \frac{\tilde{h}}{\tilde{a}^2} \frac{1}{(1 - \tilde{e}\cos u(t))^2} \right] \nonumber\\
        &= \frac{2\pi}{P(1 - e_t\cos u(t))} \frac{\tilde{h}}{\tilde{a}^2} \frac{-2\tilde{e}\sin u(t) }{(1 - \tilde{e} \cos u(t))^3} \nonumber\\
        &= \frac{2\pi }{P(1 - e_t\cos u(t))} \frac{\tilde{h}}{\tilde{a}^2}  \frac{-2\tilde{e}\sin u(t) (1 - \tilde{e} \cos u(t))}{(1 - \tilde{e} \cos u(t))^4} \label{eq:Omegadote-der}\,.
\end{align}

    To get from the first to the second line of Eq.~(\ref{eq:Omegadote-der}),
    we compute $du/dt$ by differentiating Eq.~\eqref{eq:dudt_source}. Between the fourth and fifth lines, we multiply by $1 = (1 - \tilde{e}\cos u(t))/(1 - \tilde{e}\cos u(t))$ to get an even power of $(1 - \tilde{e}\cos u(t))$ in the denominator.
To this 1PN order, we apply $(1 - \tilde{e}\cos u(t))^2 = (1 - e_t\cos u(t))(1 - e_\phi \cos u(t))$. Thus, Eq.~\eqref{eq:Omegadote-der} simplifies to
\begin{align}
    \delta \dot{\Omega}_e = - \frac{4\pi}{P} \frac{\tilde{h}}{\tilde{a}^2} \left[ \frac{\tilde{e}\sin u(t) (1 - \tilde{e} \cos u(t))}{(1 - e_t\cos u(t))^3(1 - e_\phi \cos u(t))^2} \right]\,.
\end{align}
Substituting Eqs.~\eqref{eq:PNP},~\eqref{eq:PNa}, and~\eqref{eq:PNh} yields Eq.~\eqref{omegadote1PN}.

This result has two familiar limits. 
Restricting to 0PN, i.e., taking the limit $a\rightarrow\infty$ yields
\begin{align}
\lim_{a\rightarrow\infty}\delta \dot{\Omega}_{e}(t) &=-2\frac{M}{a^3} \frac{e\sqrt{1-e^2}\sin{u(t)}}{[1-e\cos{u(t)}]^4}\,.
\end{align}
Further taking the quasicircular limit $e\rightarrow0$ yields
\begin{align}
    \lim_ {e \to 0} \delta \dot{\Omega}_e &= -2 e\frac{M}{a^3} \sin \left(\bar{\Omega}t\right),
\end{align}
which agrees with Eq. (5) of Ref.~\cite{Buonanno:2010yk}. 
\subsection{Deriving the mapping from $\theta_{\rm orb}$ to $\theta_{\rm ID}$} 
\label{input_derivation}

Again, we follow Ref.~\cite{pw} for deriving a mapping between the orbital parameters $\vec{\theta}_{\rm orb}=(a,e,\ell)$ and the initial data parameters $\vec{\theta}_{\rm ID}=(\Omega,v_r,D)$.

First, we consider the separation of the two BHs as $D$. 
In Appendix~\ref{sec:pn-app}, we introduced the Keplerian separation for an eccentric orbit as $r(\vec{\theta}_{\rm orb},\,t)$. 
We utilize this expression for the BH separation
\begin{equation}
    D(\vec{\theta}_{\rm orb}, t)      = a(1 - e\cos u(t))\,. \label{eqn:D0_formula}
\end{equation}
The $a$ and $e$ in Eq.~\eqref{eqn:D0_formula} are the Newtonian parameters for semimajor axis and eccentricity, respectively. 

Next, we consider the relative radial velocity, $v_r$, defined as $v_r = \dot{r}(t) / r(t)$, where $r(t)$ is the relative separation of the two BHs and $\dot{r}(t)$ is its time derivative. The latter is 
\begin{align}
    \dot{r}(\vec{\theta}_{\rm orb},t) &= ae \frac{du}{dt} \sin u(t) = \frac{e\sin u(t) \left[ 1 + \frac{(\eta - 9)}{2a} \right]}{ \sqrt{a}(1-e_t\cos  u(t))}\,,
\end{align}
where we have used the expression for $du/dt$ from Eq.~\eqref{eq:Omegadote-der} and substituted Eq.~\eqref{eq:PNP}. Overall,
\begin{equation}
    v_r(\vec{\theta}_{\rm orb},t)   = \frac{e \left[ 1 + \frac{(\eta - 9)}{2a} \right] \sin u(t)}{a^{3/2}(1 - e\cos u(t))(1-e_t\cos u(t))}\,,\label{inp_param_eqns-vr0}
\end{equation}

Finally, we consider the angular velocity, $\Omega (t)$.  We begin with the first line of Eq.~\eqref{eq:Omegadote-der} for the angular velocity
\begin{equation}
    \Omega(\vec{\theta}_{\rm orb},t) = \frac{d\phi}{dt}  = \frac{d\phi}{du} \frac{du}{dt}\,.
\end{equation}
Then, we substitute Eqs.~\eqref{eq:PNa} and~\eqref{eq:PNh} to achieve
\begin{equation}
    \Omega(\vec{\theta}_{\rm orb},t) = \frac{\tilde{h}}{\tilde{a}^2} \frac{1}{(1 - \tilde{e}\cos u(t))^2}\,.
\end{equation}
Further substituting the 1PN approximations of Eqs.~\eqref{eq:PNh},~\eqref{eq:PNa},~\eqref{eq:PNe} yields the final expression 
\begin{equation}
    \Omega(\vec{\theta}_{\rm orb},t)  = \frac{B\sqrt{1-e^2}}{a^{3/2}(1-e_\phi\cos u(t))(1-e_t\cos u(t))} \label{inp_param_eqns-omega0}\,,
\end{equation}
where 
\begin{equation}
    B = 1 - \frac{(3-\eta) + e^2(2\eta - 9)}{2a(1-e^2)} \,.
\end{equation}

\subsection{Spin update derivations}
\label{apx:spin_update_derivation}

Section~\ref{spin-update} provides a qualitative explanation of the spin-update procedure. 
Here, we present the relevant spin matrices, $R_{\rm BH\to in}(t=t_T)$ and $R_{0\rightarrow t_T}$. 

First, we define $R_{\rm BH\to in}(t=t_T)$, the rotation from the corotating BH frame into the inertial frame. The target spins $\vec{\chi}_{A,T}(t=t_T)$ are defined in the BH frame which might not coincide with the inertial frame at $t_T$. 
The rotation matrix $R_{\rm BH\to in}(t=t_T)$ which transforms from the BH to the inertial frame at $t_T$, is formed through the basis vectors $\hat{n} (t_T), \hat{\lambda} (t_T) $, and $\hat{L}(t_T)$, defined in the inertial frame. 
Again, we utilize the inertial frame vector notation from Ref.~\cite{Buonanno:2010yk}:
\begin{equation}
    R_{\rm BH\to in}(t=t_T)   = 
    \begin{bmatrix}
            \hat{n}_x(t_T) & \hat{\lambda}_x(t_T) & \hat{L}_x(t_T) \\
            \hat{n}_y(t_T) & \hat{\lambda}_y(t_T) & \hat{L}_y(t_T) \\
            \hat{n}_z(t_T) & \hat{\lambda}_z(t_T) & \hat{L}_z(t_T) 
    \end{bmatrix}
\end{equation}

Next, we derive the rotation that a spin vector $\vec{\chi}_{A}^{\,(i)}$ undergoes from $t=0$ to $t=t_T$, denoted $R_{0\rightarrow t_T}$. Figure~\ref{fig:spinupdateschematic} helps to visualize this rotation matrix. This rotation is applied to the spin vector at $t=0$ as follows: 
\begin{align}
    R_{0\to t_T} \vec{\chi}_{A}  &= I_3 \vec{\chi}_{A}
    + \left[\vec{\chi}_{A}^{\,(i)}(t=0) \times \vec{\chi}_{A}^{\,(i)}(t=t_T)\right] \times \vec{\chi}_{A}  \nonumber\\
    &+ \frac{\left[\vec{\chi}_{A}^{\,(i)}(t=0) \times \vec{\chi}_{A}^{\,(i)}(t=t_T)\right]^2}{1 + \vec{\chi}_{A}^{\,(i)}(t=0) \cdot \vec{\chi}_{A}^{\,(i)}(t=t_T)} \times \vec{\chi}_{A} 
\end{align}
where $I_3$ is the 3-dimensional identity matrix. The rotation angle in the inertial frame, implied by the inner and cross products of the spin vectors, corresponds to the angle between the initial and reference time spin vectors. Typical values range from a few to tens of degrees.

\bibliography{Refs}

\end{document}